\documentclass[11pt]{article}
\usepackage{graphicx}       
\newcommand{\re}{\mathrm{e}}
\newcommand{\hp}{\mathrm{h}}
\newcommand{\ri}{\mathrm{i}}

\RequirePackage{amssymb}
\usepackage{geometry} 
\begin{document}

\begin{center}
{\bf Quantum Distributions for the Plane Rotator} \\[2cm] 
Marius Grigorescu \\[3cm]  
\end{center}
\noindent
$\underline{~~~~~~~~~~~~~~~~~~~~~~~~~~~~~~~~~~~~~~~~~~~~~~~~~~~~~~~~
~~~~~~~~~~~~~~~~~~~~~~~~~~~~~~~~~~~~~~~~~~~~~~~~~~~~~~~~~~}$ \\[.3cm]
Quantum phase-space distributions (Wigner functions) for the plane 
rotator are defined using wave functions expressed in both angle  and 
angular momentum representations, with emphasis on the quantum superposition
between the Fourier dual variable and the canonically conjugate coordinate. The standard 
quantization condition for angular momentum appears as necessary for consistency. It is 
shown that at finite temperature the time dependence of the quantum wave functions may 
provide classical sound waves. Non-thermal quantum entropy is associated with localization 
along the orbit.  
\\
$\underline{~~~~~~~~~~~~~~~~~~~~~~~~~~~~~~~~~~~~~~~~~~~~~~~~~~~~~~~~~~~~~
~~~~~~~~~~~~~~~~~~~~~~~~~~~~~~~~~~~~~~~~~~~~~~~~~~~~~}$ \\

\newpage

\section{Introduction} 
The action-angle coordinates $\{ (J_i, \varphi_i), ~i=1,N/ J_i \in \mathbb{R}, \varphi_i 
\in [ - \pi, \pi ] \}$  on the phase-space $M$ arise in the description of the integrable 
Hamiltonian systems with periodic orbits \cite{am}. In these variables\footnote{In the 
standard approach $J_i$ are considered as coordinates and $\varphi_i$ as momenta. For 
systems with symmetry $J_i$ are provided by the momentum mapping and $\varphi_i$ are 
group coordinates.}  
the Hamilton function depends only on ${\bf J} \equiv \{ J_i, i = 1,N \} $, such that  
a submanifold $\Sigma_{\bf J}$ of constant ${\bf J}$ is a torus 
parameterized by $\{ \varphi_i, i = 1,N \}$.  In the old quantum mechanics $J_i$ takes 
only a discrete set of values, $J_i = n_i \hbar$, $n_i \in \mathbb{Z}$, $2 \pi  \hbar = 
\hp =6.626 \times 10^{-34}$ J$\cdot$s, such that the corresponding Lagrangian 
submanifolds $\Sigma _{{\bf n} \hbar}$ provide a partition of $M$ in  cells $b_{\bf n}$ 
of volume  $\hp^N$. However, these cells are not ordered along a complete set of local
coordinates on $M$, and in the limit $\hp \rightarrow 0$ they become singular submanifolds 
of $M$, rather than points. \\ \indent
Probability distributions of particles on $M$ may arise from thermal fluctuations, or from 
an intrinsic "quantum structure", resembling the partition in cells $b_{\bf n}$ of
finite volume. The quantum structure on $M = T^* \mathbb{R}^N$ is usually associated with
the Wigner transform \cite{wig, kirk} ${\sf f}_\psi \in {\cal F}(M)$ of the quantum "wave 
function" $\psi \in L^2 (\mathbb{R}^N)$, defined in Cartesian coordinates. For 
integrable systems the quantum distributions ${\sf f}_{\psi_n}$ provided by the eigenstates
$\psi_n$ of the Hamiltonian operator show an increased localization probability on 
$\Sigma _{{\bf n} \hbar} \subset M$ \cite{mcg}, but despite constant effort, a direct 
definition of ${\sf f}_\psi$ in terms of the action-angle variables is faced with 
difficulties. Various aspects of the problem are  presented in \cite{ka, tct, pt, bkw}.        
\\ \indent
In this work the quantum distributions for the action-angle variables are discussed on the
representative example of the plane rotator ($M = T^* S^1$). The treatment is similar to the 
one applied before to the rigid rotator \cite{cdr}, but instead of discretization here the emphasis
is on the quantum superposition between the symplectic dual (canonically conjugate) and 
Fourier dual variables. The basic elements of the formalism are presented in Section 2, 
followed in Section 3 by applications to the Wigner functions ${\sf f}_\psi$ of the plane 
rotator. Finite temperature effects, beyond the single particle coherence time, 
are discussed in Section 4. Concluding remarks are summarized in Section 5.     
 
\section{The partial Fourier transform as Hermitian operator}
Let $f(x,y)$ be a real integrable function of $x,y \in {\mathbb R}$ and $\tilde{f}(x,k)$ the 
partial Fourier transform of $f$ only with respect to $y$,
\begin{equation}
\tilde{f}(x,k) = \int dy  ~\re^{\ri k y} ~f(x,y) ~~. \label{2.1}
\end{equation}
Because $\tilde{f}(x,k)^* = \tilde{f}(x,- k)$, we may consider $\tilde{f}(x,k)$ as matrix 
element of a Hermitian operator $\hat{f}$ on $L^2({\mathbb R})$, having $x$ and $k$ as 
indices not along rows and columns, but along the diagonals \cite{cdq}. In the case of 
$M = T^* {\mathbb R}$ parameterized by the canonical variables $(q,p)$, the Fourier transform
in momentum $\tilde{f}(q,k)$ of $f \in {\cal F}(M)$ (the set of smooth functions on $M$), 
provides a matrix element  $\hat{f}_{ab} \equiv \hp^{-1} \tilde{f}((a+b)/2,(a-b)/ \hbar)$ (the 
"Weyl quantization" of  $f$) with the row and column indices $a=q+ \hbar k/2$, $b=q- \hbar k/2$  
defined using $\hbar$ as a conversion factor from $k$ to $q$-scale. Thus, if $f_1, f_2 \in 
{\cal F}(M)$ then \cite{cdq}   
\begin{equation}
(f_1, f_2) = \int_M dq dp~ f_1(q,p) f_2 (q,p)  =
\hp \int da \int db~ \hat{f}_{1ab}  \hat{f}_{2 ba}  \equiv \hp Tr( \hat{f}_1 \hat{f}_2) ~~.
 \end{equation}
The change of integration variables from $(q,p)$ to $(a,b)$ is completely formal and it does not change the physics (classical or quantum) of the observables $f_1,f_2$. However, it 
distinguishes between a pure quantum distribution ${\sf f}_\psi \in {\cal F}(M)$ and  other 
observables by reducing $\hp \hat{\sf f}_\psi$ to a projection operator,     
$\hp ( \hat{\sf f}_\psi)_{ab}= \psi_a \psi_b^*$, $\psi \in L^2(\mathbb{R})$, $\vert 
\vert \psi \vert \vert =1$. In this case the expectation value of $A \in {\cal F}(M)$ with
respect to ${\sf f}_\psi $ is 
$$
< A >_{\sf f_\psi} = ({\sf f}_\psi, A) = \hp Tr( \hat{\sf f}_\psi \hat{A}) = \langle \psi 
\vert \hat{A} \vert \psi \rangle ~~.
$$   
Similar results can be obtained using the "momentum representation", defined by the Fourier
transform in coordinate,  
\begin{equation}
\tilde{f}'(k',p) = \int dq  ~\re^{\ri k' q} ~f(q,p) ~~,
\end{equation}
such that $\hat{f}'_{b'a'} \equiv \hp^{-1} \tilde{f}'((a'-b')/\hbar,(a'+b')/ 2)$, 
with $a'=p+ \hbar k'/2$, $b'=p- \hbar k'/2$. It can be shown that if $\hp \hat{\sf f}$
is separable as $\hp \hat{\sf f}_{ab} = \psi_a \psi_b^*$, then $\hp \hat{\sf f}'$ is also 
separable, $\hp \hat{\sf f}'_{b'a'}= \psi'_{b'} (\psi'_{a'})^*$, with 
$$ \psi'_p = \frac{1}{\sqrt{2 \pi \hbar}} \int dq~ \re^{ - \ri pq/ \hbar} \psi_q ~~.$$
This result ensures that both marginal distributions are positive definite, 
$$F_\psi^{cs}(q) \equiv \int dp~ {\sf f}_\psi (q,p) = \psi_q \psi_q^*  ~~,~~
F_{\psi}^{ms}(p) \equiv \int dq ~{\sf f}_{\psi} (q,p) = \psi_p' \psi_p'^*~~,$$
and that we may consider $\psi_q$ and $\psi_p'$ as components of the same "state vector"
$\vert \psi \rangle$ in dual bases, $\vert q \rangle $ ( $\equiv \vert \hbar k \rangle $),
and $\vert p \rangle $, formally related by Fourier transform,  
$$ 
\vert p \rangle = \frac{1}{\sqrt{2 \pi \hbar}} \int dq~ \re^{ \ri pq/ \hbar} \vert q \rangle~~.
$$
It is interesting to note that the ordering of the matrix indices $a,b$ does not always 
follow the one of the variables $q,k$. Thus, for variations $\delta a >0$, $\delta b >0$ we
get also $\delta q >0$, but $\delta k >0$ only if $\delta a > \delta b$. A related aspect is
the sensitivity of ${\sf f}_\psi$ to the local inversion symmetry of $\psi$, as   
${\sf f}_\psi(q,0)$ has large negative values at the points $q_n$ where $\psi(q_n+ 
\delta q) = - \psi(q_n - \delta q)$. \\ \indent
Because $F_\psi^{cs}(q) >0$ and $F_\psi^{ms}(p) >0$, the negative values of 
${\sf f}_\psi(q,p)$ indicate that in a quantum distribution the canonical coordinates $(q,p)$ 
are not independent, but correlated by the implicit dependence of ${\sf f}_\psi$ on 
the Fourier dual  variables, $k$ or $k'$. 
A measure of these correlations is given by the function $C_\psi(q,p) = 
{\sf f}_\psi (q,p) -  F_\psi^{cs}(q) F_\psi^{ms}(p)$.   

\section{Distributions for the plane rotator}
A distribution function ${\sf f} ( \varphi, J)$ of angle ($\varphi$) and orbital angular 
momentum  ($J \equiv L_z$) on $M = T^* S^1 \simeq S^1 \times \mathbb{R}$ may describe an 
ensemble of beads on a circle, and can be regarded as a constrained distribution on 
$T^*{\mathbb R}^2$ (Appendix 1). Along the  lines of Section 2 we may also start with the 
partial Fourier transform   
\begin{equation}
\tilde{\sf f}(\varphi ,k) = \int dJ  ~\re^{\ri k J} ~{\sf f}(\varphi ,J) ~~. 
\end{equation}
To proceed towards the one-particle quantum distributions one should note that if we let 
$k \in {\mathbb R}$ and $\varphi \in [ - \pi, \pi]$ then $\alpha = \varphi + \hbar k /2$,  
$\beta = \varphi - \hbar k /2$ are not well defined as indices for a matrix element 
$\hat{\sf f}_{\alpha \beta} = \hp^{-1} \tilde{\sf f}((\alpha +\beta)/2,(\alpha - \beta)/ 
\hbar)$ of a Hermitian operator $\hat{\sf f}$ on the quantum Hilbert space ${\cal H}= 
L^2(S^1)$. Therefore, following the  example of the rigid rotator \cite{cdr}, quantum 
distributions ${\sf f}_\psi$ can be defined using a separable expression  
$\tilde{\sf f}_\psi(\varphi, k) \equiv \psi_\alpha  \psi_\beta^*$, only if the range of 
$\gamma = \hbar k$ is restricted to the first "Brillouin zone", $\gamma \in [ - \pi, \pi]$.
In this case one obtains       
\begin{equation}
{\sf f}_\psi (\varphi,J) = \frac{1}{2 \pi \hbar} \int_{- \pi}^{\pi}  d \gamma   
~\re^{- \ri \gamma J / \hbar} ~ \psi(\varphi + \frac{\gamma}{2}) \psi^*(\varphi - 
 \frac{\gamma}{2}) ~~, \label{3.2}
\end{equation}
in agreement with $V_\psi(\theta,p)$ of \cite{ka}. The overlap between two such functions is
\begin{equation}
({\sf f}_{\psi 1}, {\sf f}_{\psi 2}) = \int_{- \infty}^{\infty} dJ \int_{- \pi}^{\pi}d
\varphi ~ {\sf f}_{\psi 1} {\sf f}_{\psi 2} = \hp Tr( \hat{f}_{\psi 1} \hat{f}_{\psi 2}) =
\frac{ \vert \langle \psi_1 \vert \psi_2 \rangle \vert^2}{2 \pi \hbar}  ~~,
 \end{equation}
where $\langle \psi_1 \vert \psi_2 \rangle \equiv \oint d \varphi \psi_1^* \psi_2$
is the scalar product between  $ \psi_1$ and  $\psi_2$ as elements of ${\cal H}$. 
The marginal distributions provided by (\ref{3.2})  are
\begin{equation}
F_\psi^{cs}(\varphi) = \int dJ {\sf f}_\psi (\varphi, J) = \psi_\varphi  \psi_\varphi^*~~,  
\end{equation}
positive definite, and
\begin{equation}
F_\psi^{ms}(J) =  \oint d\varphi {\sf f}_\psi (\varphi, J) = \frac{1}{\hbar} \langle \psi 
\vert \hat{P}_J \vert \psi \rangle~~,
\end{equation}
where ($\partial_\varphi \equiv \partial / \partial \varphi$),
\begin{equation}
\hat{P}_J = \frac{1}{2 \pi} \oint d \gamma \re^{ \ri \gamma (\hat{J} - J)/ \hbar} ~~,~~
\hat{J} = - \ri \hbar \partial_\varphi ~~. 
\end{equation}
If $\psi_n (\varphi) = \re^{\ri n \varphi}/ \sqrt{2 \pi}$, $n \in {\mathbb Z}$, is an "integral" 
eigenstate\footnote{For the
harmonic oscillator Hamiltonian $H$ on $T^* \mathbb{R}$ the action variable $J = H / \omega$ 
is positive, $(\sqrt{2J}, \varphi)$ are polar coordinates on $T^* \mathbb{R}$, 
($ dp \wedge dq = dJ \wedge d \varphi$), and the eigenstates of $\hat{J}$ are real.}  of 
$\hat{J}$,  ($\hat{J} \psi_n = n \hbar \psi_n$), then  
$ \langle \psi_n   \vert \hat{P}_J \vert \psi_n  \rangle = j_0 (\pi (n - J/ \hbar))$,
$j_0 (x) = \sin (x) /x$. This shows that  $F_\psi^{ms}(J)$ is not positive definite if
$J / \hbar \in \mathbb{R}$, but if $J / \hbar \in \mathbb{Z} $ then $\hat{P}_{n \hbar}$,
$n \in \mathbb{Z}$ becomes a projection operator on $\psi_n$, $\hat{P}_{n \hbar} = \vert 
\psi_n \rangle \langle \psi_n \vert$, and  $F_\psi^{ms}(n \hbar)=  
\vert \langle \psi \vert \psi_n \rangle \vert^2  / \hbar \ge 0$. Moreover, if $J/ \hbar 
\in \mathbb{Z}$ and $\psi$ is a function of good parity, $\psi(\varphi + \pi) = \pm \psi (
\varphi)$, then the integral (\ref{3.2}) becomes intrinsic on $S^1$, namely invariant to a
change of parameter $\gamma \rightarrow \gamma + 2 \pi$.   \\ \indent
To obtain quantum distributions in the angular momentum representation the approach is 
similar, but also faced with difficulties. Because $\varphi$ has a finite range a function 
$f \in {\cal F}(M)$  can be expanded in a Fourier series,
\begin{equation}
f (\varphi,J) = \frac{1}{2 \pi \hbar} \sum_{m \in \mathbb{Z}}   
~\re^{- \ri m \varphi } ~ \tilde{f}'(m, J/ \hbar) ~~, \label{3.7}  
\end{equation}
but with $J/ \hbar \in \mathbb{R}$ and $m \in \mathbb{Z}$ we cannot take $a = J/ \hbar +m/2$   
and $b = J/ \hbar -m/2$ as indices of a matrix element. However, for ${\sf f}_\psi(\varphi,J)$ 
of the form (\ref{3.2}),  if $J / \hbar \in \mathbb{Z}$ and $\psi$ is of good parity\footnote{
This means that $m$ is an even integer. The odd values of $m$ enlarge the domain $\{a \in 
{\mathbb Z} \}$ of $\psi'_a$ by new points, $a = n +1/2, n \in  {\mathbb Z}$.}, then
$a,b  \in \mathbb{Z}$ too, and in the expansion (\ref{3.7}) $\tilde{\sf f}'_{\psi'}(m,J/\hbar)= \psi'_b (\psi'_a)^*$ with
\begin{equation}
\psi'_n = \frac{1}{\sqrt{2 \pi }} \oint d \varphi~ \re^{ - \ri n \varphi} \psi_\varphi ~~.
\end{equation}
In this representation  
\begin{equation}
F_{\psi'}^{cs}(\varphi) = \hbar \sum_{n \in {\mathbb Z}} {\sf f}_{\psi'} (\varphi, n \hbar)  
= \psi_\varphi \psi_\varphi^*~~,  
\end{equation}
(considering $\int dJ = \hbar \sum_{n = J/ \hbar}$), and
\begin{equation}
F_{\psi'}^{ms}(J) =  \oint d\varphi {\sf f}_{\psi'} (\varphi, J) = \psi'_{J/ \hbar}  
(\psi'_{J/ \hbar})^*~~.
\end{equation}
Moreover, for $f_1, f_2 \in {\cal F}(M)$ we get $(f_1, f_2) = \hp Tr'(\hat{f}_1' 
\hat{f}_2')$, where $Tr' \hat{A}' \equiv \sum_{b \in {\mathbb Z}} \hat{A}'_{bb}$ and $\hat{f}'_{ba} =
\tilde{f}'(m, J/\hbar) / \hp$, $m=a-b$, $J = \hbar(a+b)/2$ . In particular, $\hat{1}'_{ba} = \delta_{ba}$, 
 $\hat{J}'_{ba} = \hbar a \delta_{ba}$, and 
\begin{equation}
\hat{\varphi}'_{ba} = - \frac{\ri}{a-b} (-1)^{a-b} (1- \delta_{ab})~~.
\end{equation}
The angle operator $\hat{\varphi}'$ coincides with $(\hat{\phi}_{- \pi})_p$ from \cite{bp},
and corresponds to the series expansion 
$$ \varphi = - \sum_{m \ne 0} \frac{(-1)^m}{m} \sin m \varphi ~~. $$  
One should consider though $\varphi$ only as a local coordinate, because at the points of 
discontinuity $\varphi = \pm \pi$ this series contains $\pm \sin m \pi =0$, while instead of
$\pi$, as  is  $\lim_{ n \rightarrow \infty} (\pi - \varphi/n)$, the limit   
$$
\lim_{ n \rightarrow \infty} 2 \sum_{m=1}^n \frac{ \sin ( m \varphi /n)}{m}
$$
yields  $1.08949 \pi$ (the "Gibbs phenomenon"). 

\section{Coherence properties and temperature effects} 
Similarly to the case of the free particle on the ${\mathbb R}$-axis \cite{cpw}, also for the free
plane rotator the quantum distribution ${\sf f}_\psi(\varphi, J)$ is coherent, in the sense 
that if  ${\sf f}_\psi$ is a solution of the classical Liouville equation,
\begin{equation}
\partial_t {\sf f}_\psi + \frac{J}{I} \partial_\varphi {\sf f}_\psi =0~~,
\end{equation}
then $\psi$ is a solution of the Schr\"odinger equation, $\ri \hbar \partial_t \psi = 
\hat{H} \psi$, $\hat{H} = \hat{J}^2/2I$, by $I$ denoting the moment of inertia.   
\\ \indent
At a finite temperature $T$ we may consider the thermal average over quantum distributions 
of the form (\ref{3.7}), 
\begin{equation}
{\sf f}_T (\varphi,J) = \frac{1}{2 \pi \hbar} \sum_{m \in \mathbb{Z}}   
~\re^{- \ri m \varphi } ~ \tilde{\sf f}'_T(m, J/ \hbar) ~~,   \label{4.2}
\end{equation}
where $\tilde{\sf f}'_T(m, J/ \hbar) = \sum_{s \in {\cal S}} w_{s,T} \psi^s_b (\psi^s_a)^*$, $a = 
J/ \hbar +m/2$, $b = J/ \hbar -m/2$, and $w_{s,T}$ is the thermal distribution function, 
(e.g. $w_{s,T} \sim \re^{ - E_s/k_B T}$), over a set ${\cal S}$ of one-particle
states $s$ with energy $E_s$ and average angular momentum $J_s$. \\ \indent
At thermal equilibrium, during a small  single-particle coherence time $\tau$ \cite{dp}, a 
quantum wave function $\psi^s(J/\hbar + \mu,t)$, $\mu = \pm m/2$, will become     
$$
\psi^s(J/\hbar + \mu,t+ \tau) = \re^{ - \frac{\ri}{\hbar}(E_s - \mu \hbar J_s/I) \tau}  
\psi^s(J/\hbar + \mu,t)~~,
$$
such that $\tilde{\sf f}'_T$ changes into   
$$
\tilde{\sf f}'_T(m, J/ \hbar, t+ \tau) =  \sum_{s \in {\cal S}} w_{s,T} 
\re^{ - \ri \tau  m J_s/I}  \psi^s_b (\psi^s_a)^* \vert_t~~.
$$
Presuming that in the sum above we can approximate $w_{s,T}(J_s^2 - <J^2>_T) \approx 0$, 
with $<J^2>_T = \sum_s w_{s,T} J_s^2$, we get
$$
\partial_t^2 \tilde{\sf f}'_T(m, J/ \hbar, t) = \partial_\tau^2  
\tilde{\sf f}'_T(m, J/ \hbar, t+ \tau) \vert_{\tau =0} \approx - m^2 \Omega_T^2 
\tilde{\sf f}'_T(m, J/ \hbar, t)~~,
$$
where $\Omega_T^2 =  <J^2>_T /I^2$. In this approximation we find the transition expected 
in \cite{cpw},  from the complex wave functions $\psi^s$ to real classical waves, (thermal noise),   
because for a time $t >> \tau$, ${\sf f}_T$  of (\ref{4.2}) is a  solution of the classical 
wave equation $\partial_t^2 {\sf f}_T = \Omega_T^2  \partial^2_\varphi {\sf f}_T$. The 
result is independent of $\hbar$ and should hold also during a macroscopic perturbation, with the 
condition of constructive interference along the circle providing a discrete spectrum of 
"angular wavelengths".  \newpage      

\section{Concluding remarks} 
Quasiprobability distributions (Wigner functions) for the angle ($\varphi$) and orbital 
angular momentum ($J$) of the plane rotator have been defined using quantum wave functions 
expressed in both representations, $\psi_\varphi$ and $\psi'_{J/ \hbar}$. It is shown that the
"quantum" identification between the symplectic dual and the Fourier dual ($
\times  \hbar$) introduces constraints, and the integrality condition $J/ \hbar \in 
\mathbb{Z}$ appears as necessary for consistency.  \\ \indent
It is interesting to note that the Titius-Bode law for the planetary system suggests a 
constraint resembling a form of "entropy quantization", such as 
\begin{equation}
\log_2 (\frac{J}{J_G} )^3 = n~~,~~n=0,1,2,...   \label{5.1} 
\end{equation}     
where $J_G = Mc R_G$ with $R_G = 2 \gamma_G M_o/c^2$ denoting the Schwarzschild radius of 
the central body (the Sun for the planets or Jupiter for its satellites, Appendix 2). At 
the atomic scale a constraint of this type is unlikely, but for the high circular Rydberg 
levels under active investigation \cite{et1, et2}, a mixed state with a Poisson 
distribution (\ref{7.1})  could be attributed to a stage of localization along the orbit.  \\ 

\noindent
{\bf Appendix 1: Rotational coherent states } \\

Let us consider a particle of mass $M$, in uniform rotation with the angular frequency 
$\omega >0$, around the Z-axis, on a circle of radius $R$ in the XY plane. Thus, if ${\bf u}
=(u_x, u_y)$ and ${\bf v}= (v_x, v_y)$ are the position and  momentum vectors, then     
$ {\bf u}= (R \cos \varphi , R \sin \varphi)$, ${\bf v}= (- P \sin \varphi , P \cos \varphi)$,
with $P=M \omega R$ and $\varphi = \varphi_0 + \omega t$. It can be shown that a Gaussian
distribution on $T^* \mathbb{R}^2$, centered on ${\bf u}$ and ${\bf v}$, of the form   
$$
{\sf f}_{u,v} ({\bf q}, {\bf p}) = \frac{1}{\pi^2  \hbar^2} \re^{- ({\bf q} - {\bf  u})^2/ a
- ({\bf p}- {\bf v})^2/ b} ~~,~~a = \hbar^2/b = \hbar/M \omega 
$$
can be obtained by a standard Wigner transform of the rotational coherent state ("symmetry
breaking vacuum"),  
\begin{equation}
\vert z \rangle= \re^{z \hat{b}^\dagger_u- z^* \hat{b}_u} \vert 0 \rangle ~~,~~ 
\end{equation}
where $z= \sqrt{J/ \hbar} \re^{ - \ri \varphi}$, $J =M \omega R^2$, and $\hat{b}^\dagger_u =
(\hat{b}^\dagger_x + \ri \hat{b}^\dagger_y) / \sqrt{2}$, with 
$ \hat{b}_q = \sqrt{ M \omega/ 2 \hbar} (\hat{q} + \ri \hat{p}_q /  M \omega )$,
$\hat{b}_q \vert 0 \rangle =0$, $q=x,y$.    
Moreover, the average of ${\sf f}_{u,v}$ over $\varphi \in [-\pi, \pi]$ at constant $J$ is
the Wigner transform of the density operator   
\begin{equation}
\hat{\rho}_w = \sum_{n=0}^\infty w_n \vert n \rangle \langle n \vert~~,~~ \vert n \rangle =
\frac{1}{\sqrt{n!}} (\hat{b}^\dagger_u)^n \vert 0 \rangle~~,  \label{7.1}
\end{equation} 
expressed by a Poisson (non-thermal) distribution $w_n= \re^{- J/ \hbar} (J/ \hbar)^n/n!$
of quantum entropy
$S_q = - \sum_n w_n \ln w_n$ \cite{cdq}, over the eigenstates $\vert n \rangle$ of the 
angular momentum operator $\hat{L}_z = \hbar (\hat{b}^\dagger_u \hat{b}_u - 
\hat{b}^\dagger_d \hat{b}_d)$, $\hat{b}^\dagger_d =(\hat{b}^\dagger_x - \ri 
\hat{b}^\dagger_y) / \sqrt{2}$. Worth noting, a Gaussian distribution on $\mathbb{R}$ with 
the same mean and variance as $\{ w_n \}$ has the entropy $S_J=[1+ \ln (2 \pi J /\hbar)]/2$. \\

\noindent
 {\bf Appendix 2: Comparison with astronomical data} \\

The condition (\ref{5.1}) and the third law of J. Kepler (May 15 1618) yield for the  
n'th circular orbit the radius $r_n = R_G 2^{1+ 2n/3}$, $n=0,1,2,...$ . For the Sun 
$R_G= 2.95$ km, and with $n \ge 35$ the calculated values $r_n$ are close to the 
astronomical data for all planets (e.g. the Earth's orbit has an average radius of 
$0.95 r_{37}$), excepting Jupiter, whose average orbital radius is $0.96( r_{40}+r_{41})/2$. \\ \indent   
In the case of Jupiter $R_G =2.82$ m, and for $n=39,40,41,42$ (Table 1), we get values\footnote{$r_{35}=59 609$ km 
corresponds to the "internal" orbit close to the 1 bar surface of radius $71 492$ km.}
($r_n$)  close to the ones  observed ($r_{obs}$) for its largest satellites: Io, Europa, Ganimede, Callisto, discovered by 
G. Galilei in 1610.  \\ \indent 
Apart from the scale factor $R_G$, a peculiar common trait for these systems is the weighted 
average $\bar{n}$ of the orbital number,
\begin{equation}
\bar{n} = \frac{ \sum_n n M_n }{\sum_n M_n } ~~,
\end{equation}
where $M_n$ denotes the observed mass of the body assigned to the radius $r_n$. Thus, one obtains 
$\bar{n} =41$ for the planetary system (considering  $M_{40} = M_{41} = 0.5 M_{Jupiter}$), and
$\bar{n} = 40.7$ for the satellites of Table 1.  \\ \indent
In a historical perspective, the Titius-Bode law (1766-72) was anticipated by Kepler's 
geometric model of the solar system, (in {\it Mysterium Cosmographicum}, T\"ubingen, 1596), where 
the intervals between the planetary orbits are not arbitrary, but determined by 
"perfect" solids. 
Less known is that in fact, as shown in \cite{gsa}, the standard 
quantization  of angular momenta can be related to these solids. \\

\noindent
{\bf Table 1.} Comparison between the observed orbital radius ($r_{obs}$) of the Jupiter 
satellites and the calculated value ($r_n$).   \\

\begin{tabular}{|c|c|c|c|c|c|c|c|}
\hline
Satellite/n  &   Io/39   &  Eu/40  &  Ga/41 & Ca/42  \\ \hline
 $r_{obs}$ ($10^3$ km) & 421.6   & 670.8   & 1070  & 1882 \\ \hline
 $r_n$     ($10^3$ km) & 378.5  & 600.8   & 953.7  & 1514 \\ \hline
 $r_{obs} / r_n$ & 1.11   & 1.12   & 1.12  & 1.24 \\ \hline
\end{tabular}

\end{document}